%% file: main.tex
\documentclass{article}

\usepackage{microtype}
\usepackage{graphicx}
\usepackage{caption}
\usepackage{subcaption}
\usepackage{booktabs} 

\usepackage{hyperref}

\usepackage[preprint]{icml2026}

\usepackage{amsmath}
\usepackage{amssymb}
\usepackage{mathtools}
\usepackage{amsthm}
\usepackage{nicefrac}
\usepackage{subfiles} 
\usepackage{minitoc}

\usepackage[capitalize,noabbrev]{cleveref}

\theoremstyle{plain}

\theoremstyle{definition}

\theoremstyle{remark}

\usepackage[textsize=tiny]{todonotes}

\icmltitlerunning{JEDI: Jointly Embedded Inference of Neural Dynamics}

\usepackage{stmaryrd}
\usepackage{url}
\usepackage{graphicx} 
\usepackage{xcolor}
\usepackage{xspace}
\usepackage[dvipsnames]{xcolor}
\usepackage[utf8]{inputenc}
\usepackage[inline]{enumitem}
\usepackage{enumitem}
\usepackage{stmaryrd}
\usepackage{bibentry}
\usepackage{booktabs}
\usepackage{siunitx}
\usepackage{comment}
\usepackage{subcaption}
\usepackage{soul}
\usepackage{array}
\usepackage{bbm}
\usepackage{placeins}
\usepackage{makecell}
\usepackage{nicefrac}
\usepackage{graphbox}
\usepackage{textcmds}
\usepackage{wrapfig}

\usepackage[utf8]{inputenc} 
\usepackage[T1]{fontenc}    
\usepackage[export]{adjustbox}

\newcommand{\beginsupplement}{%
        \newpage
        \setcounter{table}{0}
        \renewcommand{\thetable}{S\arabic{table}}%
        \setcounter{figure}{0}
        \renewcommand{\thefigure}{S\arabic{figure}}%
        \appendix
}

\begin{document}

\twocolumn[
  \icmltitle{JEDI: Jointly Embedded Inference of Neural Dynamics} 



  \icmlsetsymbol{equal}{*}
  \icmlsetsymbol{dagger}{\dagger}

  \begin{icmlauthorlist}
    \icmlauthor{Anirudh Jamkhandi}{yyy,comp}
    \icmlauthor{Ali Korojy}{yyy,comp}
    \icmlauthor{Olivier Codol}{yyy,comp}
    \icmlauthor{Guillaume Lajoie}{equal,yyy,comp}
    \icmlauthor{Matthew G. Perich}{equal,yyy,comp}

  \end{icmlauthorlist}

  \icmlaffiliation{yyy}{Mila - Quebec AI Institute}
  \icmlaffiliation{comp}{Université de Montréal}

  \icmlcorrespondingauthor{Anirudh Jamkhandi}{anirudh.jamkhandi@umontreal.ca}
  \icmlcorrespondingauthor{Matthew G. Perich}{matthew.perich@umontreal.ca}

  \icmlkeywords{Neural Dynamics, RNN, Hypernetworks, Neuroscience}

  \vskip 0.3in
]



 \printAffiliationsAndNotice{\icmlEqualContribution}

\input{sections/abstract}

\input{sections/introduction}

\input{sections/methodology}

\input{sections/experiments}
\input{sections/related_work}
\input{sections/discussion}


\bibliography{main}
\bibliographystyle{icml2026}


\newpage
\beginsupplement
\newpage
\onecolumn
\input{sections/appendix.tex}

\end{document}

%% file: sections/abstract.tex
\begin{abstract}

Animal brains flexibly and efficiently achieve many behavioral tasks with a single neural network. A core goal in modern neuroscience is to map the mechanisms of the brain’s flexibility onto the dynamics underlying neural populations. However, identifying task-specific dynamical rules from limited, noisy, and high-dimensional experimental neural recordings remains a major challenge, as experimental data often provide only partial access to brain states and dynamical mechanisms. While recurrent neural networks (RNNs) directly constrained neural data have been effective in inferring underlying dynamical mechanisms, they are typically limited to single-task domains and struggle to generalize across behavioral conditions. Here, we introduce JEDI, a hierarchical model that captures neural dynamics across tasks and contexts by learning a shared embedding space over RNN weights. This model recapitulates individual samples of neural dynamics while scaling to arbitrarily large and complex datasets, uncovering shared structure across conditions in a single, unified model. Using simulated RNN datasets, we demonstrate that JEDI accurately learns robust, generalizable, condition-specific embeddings. By reverse-engineering the weights learned by JEDI, we show that it recovers ground truth fixed point structures and unveils key features of the underlying neural dynamics in the eigenspectra. Finally, we apply JEDI to motor cortex recordings during monkey reaching to extract  mechanistic insight into the neural dynamics of motor control. Our work shows that joint learning of contextual embeddings and recurrent weights provides scalable and generalizable inference of brain dynamics from recordings alone.
\end{abstract}

%% file: sections/introduction.tex
\section{Introduction}
Brains are complex and nonlinear dynamical systems composed of a network of highly specialized neural circuits. The neural computations that govern complex behaviors (e.g., playing guitar or writing machine learning papers) ultimately arise from the time-varying interactions of neural populations distributed across the brain \citep{duncker2021dynamics}. With modern advances in recording technologies, experimentalists can record the activity of ever-larger populations of neurons, providing an unprecedented window into neural computation. A critical question then arises: How can we obtain mechanistic insights into the neural dynamics underlying these behaviorally-relevant neural computations?

The structure of neural interactions plays a central role in systems neuroscience, where the patterns and strength of connectivity between neurons shape the resulting dynamics and computation \citep{liu2024connectivity,schuessler2024aligned,ostojic2024computational}. 
Analysis of these interaction weights can reveal how connectivity constrains and enables flexible computations \citep{braun2022exact,raman2021frozen,canatar2021spectral}, but these weights are not readily accessible in typical experimental settings. 

Over the past couple of decades, recurrent neural networks (RNNs)—which incorporate many canonical features of neural circuits such as recurrence and nonlinear interactions—have emerged as a powerful model for studying neural population activity \citep{Sussillo2009,Hess2023,Dinc2023,Valente2022,Perich2020,durstewitz2023reconstructing}. RNNs trained to perform neuroscience-inspired tasks can be compared to animals performing similar tasks, allowing for indirect exploration of mechanisms in the biological brain. However, a growing class of data-constrained RNNs (dRNNs) trained to recapitulate neural recordings provide a complementary approach to make direct inference of brain computations.

Ultimately, most existing dRNNs typically assume a fixed set of weights corresponding to one dynamical system for the observed data (e.g., each behavior). In contrast, biological networks flexibly operate in different dynamical regimes even within a single task \citep{huang2024measuring, turner2021charting}. Thus, dRNN models should have similar dynamical flexibility to accurately capture neural computations necessary for complex and evolving behavioral demands. Fixed weight dRNN models lack the flexibility to account for changes in behavioral tasks or contexts. This limitation underscores the need for dRNN approaches that can accommodate context-dependent variations in neural computations and generalize robustly across behavioral tasks and conditions. 

To address this limitation, we introduce JEDI - Jointly Embedded Dynamics Inference from neural data. JEDI uses a hierarchical, hypernetwork-based framework to infer neural dynamics directly from time series of neural population recordings (Fig.~\ref{fig:overall_figure}). It leverages a shared embedding space to learn the parameters of arbitrary dRNNs that recapitulate the time series of recorded neural data (e.g. one trial of a behavioral task) using only a contextual input to the embedding. Learning in this shared model allows the model to capture common dynamics across datasets, tasks, or contexts, while providing flexibility for sample-specific dynamical features. Further, by jointly learning RNN weights and embeddings, we can infer interpretable link between weight structure and neural computations.

Our contributions can be summarized as follows:
\begin{itemize}
    \item We propose a novel hypernetwork framework to model complex neural data from varying behavioral conditions and contexts in a single unified model.
    \item Using the learned task embeddings, we demonstrate accurate classification of dynamical regimes and generalization to unseen samples from learned tasks.
    \item We show robust inference of dynamical rules from RNN weights through the eigenspectra, Lyapunov exponents, and fixed point structure of learned weights.
\end{itemize}

%% file: sections/methodology.tex
\section{Methods}

\subsection{Low-rank RNNs}

RNNs are universal approximators for dynamical systems, whose internal computations can be reverse engineered through the recurrent weights. We use rate-based recurrent network of N units, in which the variable $r$ follows the dynamics:

\begin{align}
\label{eq:RNN_cont}
\tau \frac{d\mathbf{r}}{dt}=-\mathbf{r}(t)+\mathbf{J}\phi(\mathbf{r}(t)) + \xi(t),
\end{align}

\begin{figure}[!ht]
\includegraphics[width=\linewidth]{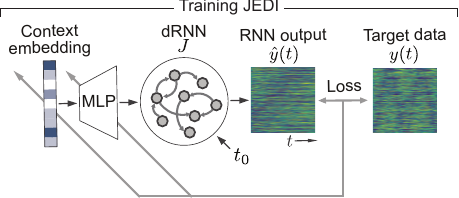}
\caption{JEDI leverages a hypernetwork-based framework to flexibly generate RNN weights based on contextual inputs, all learned directly from a loss computed against the RNN output and time series neural recordings.}
\label{fig:overall_figure}
\end{figure}

with neuron activity $\mathbf{r}(t)\in \mathbb{R}^N$, time-constant $\tau \in \mathbb{R}_{>0}$, recurrent weights $\mathbf{J}\in\mathbb{R}^{N\times N}$, element-wise nonlinearity $\phi(x)=\textit{tanh}(x)$, private white noise $\xi(t)\in R^{N}$ provided to each neuron.

 The possible combinations of all $\mathbf{J}$ that can reproduce a given sample of neural activity is potentially large. To reduce this solution degeneracy \citep{huang2024measuring}, we constrain $\mathbf{J}$ with a low-rank penalty, compressing the recurrent weight matrix into a few dominant modes. Hence, we are interested in the case where the recurrent weight matrix has rank $R\le N$, i.e., it can be written as $\mathbf{J}={MN}^{\mathsf{T}}$, with $M,N \in \mathbb{R}^{N\times R}$ \citep{Mastrogiuseppe2018,Schuessler2020,Beiran2021,Dubreuil2022}. In the case where the recurrent weights are unconstrained we will call the model \qq{full rank}.

\subsection{Context-informed Hypernetwork}

Drawing inspiration from the top-down modulation of circuits in the neocortex, we introduce conditioning through a hypernetwork \citep{ha2016hypernetworks}—a secondary network that dynamically generates the parameters of a primary network (here, a dRNN). The hypernetwork acts as a controller that receives a specific context signal (like a task instruction or a sensory goal) and uses it to dynamically reconfigure the primary network. Instead of the primary network having a single, static weights, it is conditioned to adapt its internal dynamics on the fly. Our implementation uses a feedforward layer that learns context embeddings to flexibly parameterize the downstream RNN responsible for reconstructing the dynamics of neural recordings. 
\begin{align}\label{eq:HNet}
J = f_h(c)\end{align}
where ${f_h}$ is a non-linear function parameterized by $\Theta_h$ and $c\in \mathbb{R}^{E}$ is the context vector, $E$ is the size of the context vector. Hence, equation (1) can be reformulated as : 
\begin{align}\tau \frac{dr}{dt}=-r(t)+f_{h}(c)\phi(r(t)) + \xi(t),
\label{eqn:hnet_recurrent}
\end{align}
From Equation (\ref{eqn:hnet_recurrent}) it is clear that context vector can significantly affect the overall dynamics for appropriate choices of $f_h(\cdot)$. Modeling choices for the context $c$ can influence the degree of granularity of the hypernetwork. Indeed, $c$ could be made to encode any variable, e.g. the identity of a specific trial, subject, or behavioral task. Rather than learn recurrent weights for each trial of neural activity data separately, we generate the trial-specific weights from a single, learned hypernetwork that shares efficiently information across trials. When the recurrent weights are constrained to have low rank structure, the output of the hypernetwork are the matrices $\mathbf{M},\mathbf{N}$. In our experiments, we initialize a new context $c$ with zero entries for each trial.

\subsection{Training Objective}

The goal is to estimate the recurrent weights to reproduce the recorded neural population activity through minimization of a loss function. We use back-propagation through time to minimize the squared difference between the predicted and ground truth trajectories. The mean squared error (MSE) loss on each trial $j$ can be expressed as: 
\begin{equation}
\mathcal{L}(\Theta_h, c) = 
\sum_{i=1}^{N} \sum_{t=1}^{T} 
\left( \phi\!\left(r_i^{(c)}(t)\right) - \phi\!\left(\hat{r}_i^{(c)}(t)\right) \right)^2
\end{equation}
where $r^{(c)}(t))$ represents the target trajectory in condition $c$, for neuron i and timestep t and $\hat{r}_i^{(c)}(t)$ the corresponding predicted trajectory of the model. Similar to \citet{von2019continual}, we treat the context vector as a differentiable, deterministic parameter that can be optimized alongside $\Theta_h$. At each learning step, the current context embedding $\mathbf{c}^{(j)}$ is updated along with the hypernetwork weights $\Theta_h$ to minimize the reconstruction loss. After training, the resulting context is stored and added to the collection $\{\mathbf{c}^{(j)}\}$. The nature of this collection will depend on the granularity chosen for the context embedding. In the following work, this collection represents trials from a set of tasks, dynamical contexts, or behavioral conditions, but for other applications could comprise experimental subjects or even brain regions.

%% file: sections/experiments.tex
\section{Empirical Results}
\begin{figure*}[!ht]
\includegraphics[width=\linewidth]{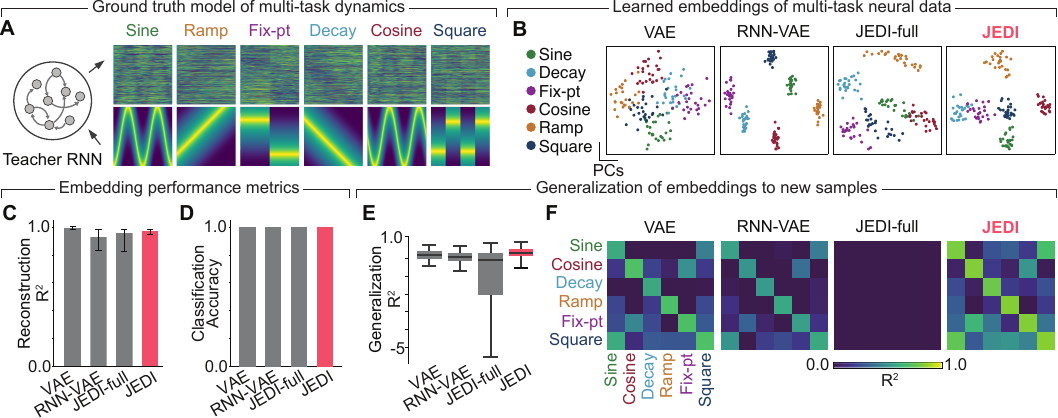}
\caption{Quantifying the quality of the embeddings.
A) Synthetic multi-task data generation setup.
B) 2D PCA visualization of context embeddings. Each point corresponds to a trial, color-coded by input signal type.
C) Training reconstruction accuracy $R^2$ across different methods.
D) Accuracy of task classification from the learned embeddings. 
E) Generalization accuracy $R^2$ from the center of the learned embedding for each task
F) Confusion matrix of generalization $R^2$ scores, expanding the results in Panel E. Rows represent training tasks, and columns indicate test tasks.}
\label{fig:synth_results}
\end{figure*}
In the following sections, we validate JEDI and demonstrate its ability to infer the dynamical structure of recorded neural population activity. Since JEDI jointly learns contextual embeddings and neural dynamics, we designed a series of experiments to explore these two properties. We first use a ground truth dataset modeling multi-task scenarios to show JEDI learns robust and generalizable embeddings. We use this experiment to explore how architectural decisions, such as low-rank weight structure, impacts model performance. We then leverage two further datasets to show that JEDI can accurately infer dynamical features by spectral analysis of the eigenvalues and fixed points analysis. Lastly, we demonstrate JEDI's efficacy in modeling real neural recordings.

\subsection{Teacher Setup for generating synthetic data}

To evaluate the performance of the model, we studied a teacher-student paradigm \citep{saad1995exact, seung1992statistical, beiran2024prediction} where a teacher network is used as a proxy for a neural system with known recurrent connectivity. The student (our model) is trained to mimic the teacher. We generated synthetic data from a chaotic teacher RNN with pre-defined recurrent connectivity and input structure (Fig.~\ref{fig:synth_results}.A). The RNN was composed of N=200 neurons, in which the activity (firing rate) $h_i(t)$ of the neuron i evolved according to the dynamical update:
\begin{align}
\tau \frac{dh}{dt}=-h(t)+gJ\phi(h(t)) + W_{ext}U(t),
\label{eq:TeacherRNN_cont}
\end{align}
where $h(t)\in\mathbb{R}^N$ is the RNN state at time $t$, $\phi$ is nonlinear activation function, $\Delta t$ is the simulation timestep, $\tau$ is the neural time constant, and $u(t)\in\mathbb{R}^N$ is time varying external input. $\mathbf{J}\in\mathbb{R}^{N\times N}$ is the recurrent connectivity matrix sampled from a Gaussian distribution $\mathcal{N}(0, \sigma^2)$. To capture the low-dimensional structure inherent in neural activity, we impose a low-rank constraint on the connectivity matrix $J$ \citep{perich2025neural} by initializing it with rank $R=5$. The gain parameter $g$ determines the strength of the recurrent connections, and thus whether ($g>1$) or not ($g<1$) the network produces spontaneous activity with non-trivial dynamics \citep{rajan2010stimulus}. We set $g=1.8$ to produce chaotic dynamics shaped by the external inputs. $W_{ext}\in R^{N\times N}$ is input matrix that maps input signals to the recurrent matrix.

To probe JEDI's ability to capture diverse dynamical regimes in a single model, we drove the teacher RNN with a range of external input signals $u(t)$ that shaped the temporal evolution of the RNN (Fig.~\ref{fig:synth_results}.A). Inputs included oscillatory (sine, cosine, square), ramping, decaying, and moving fixed-point (step-like) patterns. These inputs were directly applied to a subset of neurons in the chaotic RNN (50\% of the population), selected via an input weight matrix $W_{\text{ext}} \in \mathbb{R}^{N}$. We refer to the input types as tasks, and simulated the variability across trials by repeating each input from different random initial RNN states. We simulated the activity of $i^{th}$ neuron for 2 seconds following a 0.1 second burn-in period. The resulting activity serves as ground-truth trials of a known dynamical system, providing a testbed for assessing the generalization and interpretability of our model. 
 
We trained JEDI models to recapitulate the teacher RNN's activity across all tasks. For our first experiment, we aimed to demonstrate robust and generalizable learned embeddings capturing relationships in neural dynamics across tasks. We compared JEDI against three common approaches to learn low-dimensional embeddings of neural dynamics across trials and tasks. We trained Variational Autoencoders—both feedforward (VAE) and recurrent (RNN-VAE) variants—to reconstruct neural activity on each trial from a low-dimensional embedding space. Note that these alternative methods are intended to contextualize the utility and performance of JEDI's embeddings, but are not fully comparable to JEDI since they do not provide the mechanistic insight into neural dynamics through interaction weights.

\subsection{JEDI learns generalizable embeddings across tasks}

We trained JEDI to recapitulate the activity generated by the teacher RNN using two model configurations: one generating full rank RNN weights, and one constrained to produce RNN weights with rank $r$ (set to 5 to match the ground truth data). We configured the hypernetworks as 3-layer feedforward networks (MLPs) and compared performance of these JEDI models against the VAE and RNN-VAE methods. Using coefficient of determination ($R^2$) as our metric, we found that all models achieve high reconstruction accuracy of the teacher RNN activity (Fig.~\ref{fig:synth_results}.C). Notably, JEDI and VAE gave more consistently good fits (smaller variance across trials) than JEDI-full and RNN-VAE. 

A good contextual embedding should learn separable and interpretable structure across tasks. To evaluate if the embeddings learned the multi-task structure, we trained a Naive Bayes classifier on the learned representations and evaluated its accuracy on held-out trials. All models achieved near-perfect classification accuracy, (Fig.~\ref{fig:synth_results}.D), indicating that the dynamical regimes induced by different input types are readily separable in latent space. However, a more stringent test is whether these embeddings are generalizable and robust. 

We evaluated across-task generalization by sampling the mean embedding for each task from training data and used this embedding to generate predictions on held-out test data across all tasks. A model that captures only task-specific identity will have high values along the diagonal (generalization within task), whereas a model that uncovers shared structure across tasks will show off-diagonal generalization. While some tasks were dynamically distinct and should have no generalization (e.g. ramp vs sine), others have common structure which could be leveraged by the embedding (e.g. the oscillatory tasks). JEDI outperformed other models in across-task generalization (Fig.~\ref{fig:synth_results}.E). For example, embeddings derived from sine generalize to cosine, and embeddings from ramp transfer to decay(Fig.~\ref{fig:synth_results}.F). Interestingly, JEDI-full performed substantially worse, indicating that structural low-rank constraints in the RNN weight outputs of JEDI enable learning generalizable representations in the embedding space.

\begin{figure}[!t]
\includegraphics[width=\linewidth]{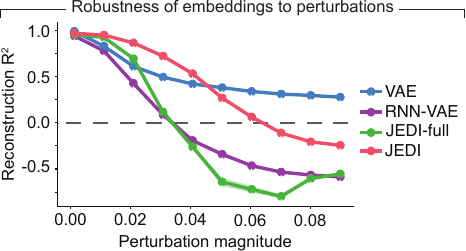}
\caption{Impact of embedding noise on model performance, comparing reconstruction $R^2$ as increasing noise is applied to the embeddings. The added noise was scaled according to the standard deviations of embeddings for each model}
\label{fig:syn_perturb}
\end{figure}

We next tested the robustness of the embeddings by adding gaussian noise of increasing variance to the learned values for each trial and decoding the resulting activity. Because JEDI’s embeddings were not constrained to Gaussian distributions like the VAEs, we scaled the perturbation magnitude according to the specific variance of each model's task embeddings. JEDI-full and RNN-VAE exhibit sharp performance declines (Fig.~\ref{fig:syn_perturb}) in the presence of the perturbations. While VAEs retain some performance at very high noise levels, JEDI outperforms all other models during low-to-moderate perturbations, indicating that the model learned robust, structured representations in the embedding.

\subsection{JEDI uncovers ground truth spectral properties of neural dynamics}
\begin{figure}[!hb]
\includegraphics[width=\linewidth]{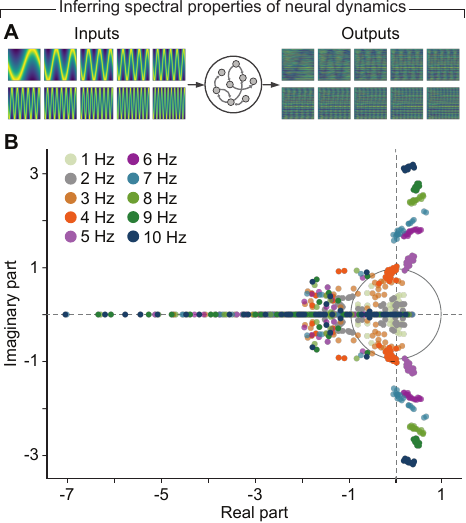}
\caption{
A) We drove the chaotic Teacher RNN with sinusoidal inputs at different frequencies (1–10 Hz).
B) The eigen spectra of the weights inferred with JEDI exhibit a characteristic expansion along the imaginary axis as input frequency increases.}
\label{fig:freq_eigen}
\end{figure}

We next explore JEDI's ability to infer dynamical properties of neural data by reverse-engineering the RNN weights. We devised a variant of the previous experiment where the teacher RNN was driven by sinusoidal inputs of increasing frequencies (Fig.~\ref{fig:freq_eigen}.A). After training on multiple samples of each frequency, we analyzed the eigen spectrum of the learned JEDI RNN weights. The imaginary part of the eigenvalues indicate rotational velocity, while the real part of eigenvalues indicate stability in associated eigen directions: smaller than zero corresponds to stability and greater than zero, instability.

We observed that eigenvalues formed distinct clusters within the complex plane based on the input frequency of the associated trials. Crucially, as the input frequency increased, the eigenspectra exhibited a corresponding expansion along the imaginary axis, directly capturing the higher oscillatory content inherent in the driving signal. In contrast, the real components remained relatively invariant, suggesting that JEDI selectively adapts the rotational dynamics of the network to match the data's spectral properties without compromising global stability. Notably, this precise spectral alignment was absent in models trained via standard full-rank initialization (see Appendix ~\ref{fig:appen_full_multi_fre}), which failed to recover the underlying dynamical structure. Furthermore, while we focus here on controlled sinusoidal inputs, JEDI demonstrates similar spectral consistency in more complex settings; for a comparison with the original eigen spectra derived from multi-task learning paradigms (see Appendix ~\ref{fig:appen_full_synth}).

\begin{figure}[!ht]
\includegraphics[width=\linewidth]{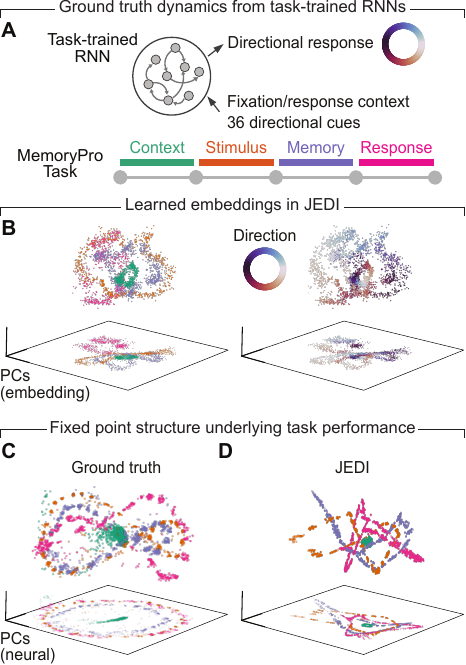}
\caption{JEDI identifies fixed point structure in task-trained networks.
A) We fit JEDI to a network trained to perform the MemoryPro task. This task has four contextual trial phases.
B) Context embedding learned by JEDI colored by trial phase (left) and response direction (right).
C) Fixed point structure for the four trial phases for the ground-truth task-trained network.
D) Fixed point structure inferred by JEDI.}
\label{fig:ctd_fp}
\end{figure}

\subsection{JEDI can recover ground truth fixed point structure present in the neural dynamics}

We explored the ability of JEDI to infer mechanisms of neural computation using the framework proposed by \citet{yang2019task}, who trained RNNs to perform a range of neuroscience-like tasks. Following the protocol of \citet{versteeg2025computation}, we used the multi-task trained RNN to simulate trials for the MemoryPro task across four distinct periods: context, stimulus, memory, and response, see (Fig.~\ref{fig:ctd_fp}.A). The fixed-point structure of the MemoryPro task was previously characterized by \citet{driscoll2024flexible}, providing a clear ground truth target of dynamical mechanisms to infer by JEDI (see Appendix ~\ref{appx:ctd_task_des} for comprehensive details).

After training on the large number of trials and contexts, JEDI yielded a low reconstruction loss and high $R^2$(0.94). We first examined whether the learned context embeddings preserved the functional organization of the task. We found that JEDI’s embeddings capture the task’s logical structure. Specifically, we found distinct, phase-specific rings organized according to stimulus direction (Fig.~\ref{fig:ctd_fp}.B left), confirming that JEDI recovers the geometric relationships between input conditions (Fig.~\ref{fig:ctd_fp}.B right).

Next, to reverse-engineer the trial-specific computations, we identified slow-moving points within the model's hidden state space—commonly referred to as fixed points \citep{sussillo2013opening}—derived from JEDI’s trained weights. These fixed points were projected into separate low-dimensional principal component (PC) spaces and visualized collectively for analysis. We observed that the fixed-point structure extracted by JEDI closely mirrored the shape of the original task-trained RNN (Fig.~\ref{fig:ctd_fp}.C\&D). RNN states evolved from a single, central fixed point that cued the context towards geometrically related, target-specific fixed points, forming an approximate ring attractor. This is consistent with the fixed-point characteristics previously reported in \citet{driscoll2024flexible} for RNNs performing this task. This alignment with established structures confirms that JEDI’s embeddings and weights effectively capture meaningful features of computations underlying neural activity.

\begin{figure*}[!ht]
\includegraphics[width=\linewidth]{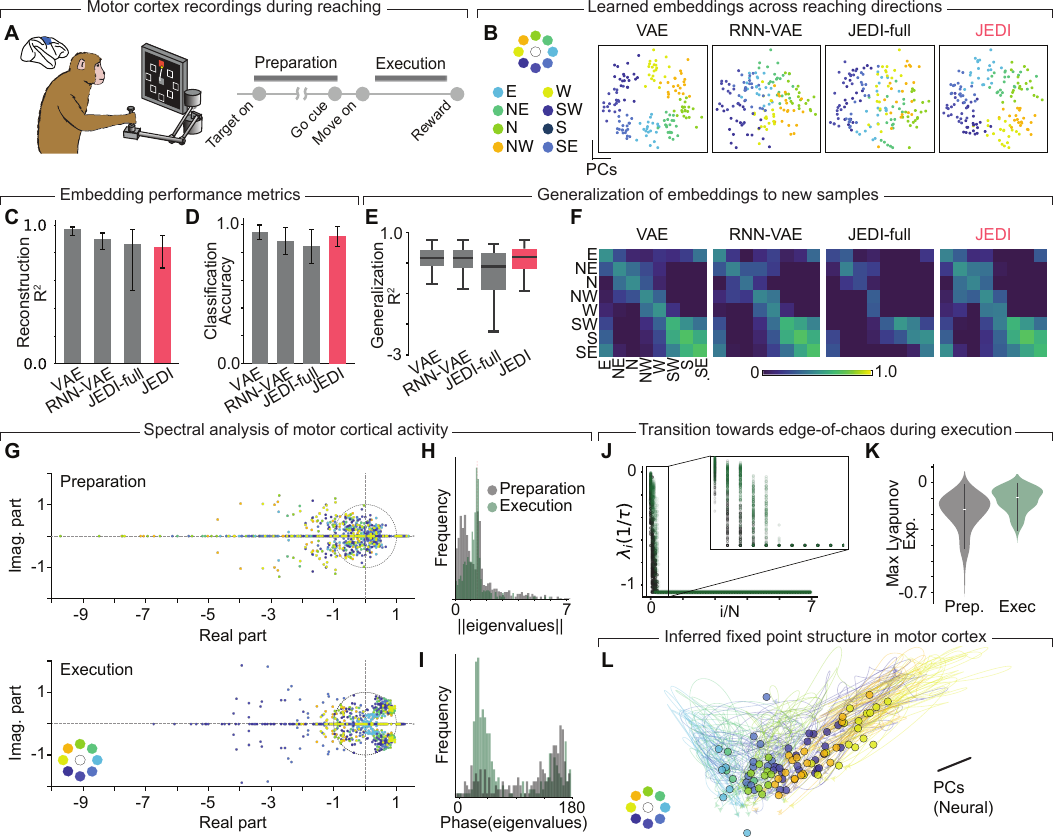}
\caption{JEDI applied to monkey motor cortex data during reaching.
A) We apply JEDI to recordings of the motor and premotor cortex during a center-out reaching task. Schematic adapted from \citet{gallego2020long}.
B) 2D PCA projections of the learned embeddings for each model. Each point corresponds to a single trial, color-coded by reach direction. 
C) Reconstruction $R^2$ across all trials for each model.
D) Classification accuracy for reach direction from each learned embedding.
E) Generalization $R^2$ using the center of each embedding cluster for the different models.
F) Confusion matrix for generalization across different contexts for each model, further visualizing the results in Panel E. 
G) Eigenspectra of learned JEDI weights during preparation (top) and movement (bottom) colored by the 8 reach directions. During movement, eigenvalues clustered close to the unit circle, indicating dynamics approaching the edge of stability.
H) Quantification of eigenspectra changes by the magnitude of each eigenvalue for preparation (gray) and execution (green).
I) Quantification of eigenspectra changes by the phase of each eigenvalue.
J) Lyapunov exponents across all learned weights for preparation and execution.
K) Distributions of maximum Lyapunov exponents across all learned trials for preparation and execution.
L) Stable fixed points during execution inferred by JEDI colored by reach directions. Single-trial neural trajectories for each corresponding reach is also plotted.}
\label{fig:embacc_monkey}
\end{figure*}

\subsection{JEDI flexibly models population recordings from monkey motor cortex}

We then applied JEDI to infer dynamical properties from real neural recordings. We fit JEDI models on motor cortical recordings from macaque monkeys performing a center-out reaching task \citep{perich2018neural}. Neural population activity was recorded simultaneously from the primary motor and premotor cortex with two electrode arrays. The monkey was trained to perform an instructed delay task, which involved a movement preparation phase followed by a go cue instructing the monkey to reach to one of eight targets (movement phase) (see Fig.~\ref{fig:embacc_monkey}.A). After training,  JEDI learned a clear ring-like structure in the embedding corresponding to the eight reach directions (Fig.~\ref{fig:embacc_monkey}.B), with comparable performance in fit quality, classification accuracy, and generalization as the other, embedding-specific methods (Fig.~\ref{fig:embacc_monkey}.C-E).

\subsection{Spectral analysis reveals reorganization of dynamics during movement in monkey reaching}
While interpretable embeddings are a useful component of our model that helps contextualize our learned dynamcis, the ultimate goal of JEDI is to directly infer dynamical mechanisms from neural data. To explore this, we compared movement preparation (a covert process without explicit motor output) to movement execution (an overt process that moves the arm). These two processes have been shown to involve distinct neural computations \citep{elsayed2016reorganization}, and we hypothesized that they should consequently have distinct dynamical properties which JEDI can uncover.

Spectral analysis of the trained recurrent matrices generated by JEDI (but not JEDI-full, see Appendix  \ref{fig:appen_full_motor}) revealed clear changes in dynamical structure across the two behavioral phases. During preparation, the eigenvalue distributions were largely clustered within the unit circle except for prominent excursions largely along the real axis (Fig.~\ref{fig:embacc_monkey}.G top), consistent with the need to generate ramping dynamics that take the brain from a quiescent state to one ready to produce behavior. During movement, however, groups of eigenvalues emerged that were tightly clustered along the unit circle near the zone of marginal stability (Fig.~\ref{fig:embacc_monkey}.G bottom), a desirable regime for efficient neural computation \citep{legenstein2007edge}. Reassuringly, these dynamics were identical across all trials, consistent with the assumption that the underlying dynamical system driving movement does not change for different reach directions. 

We further explored whether execution-related dynamics could represent a shift towards criticality (edge of stability). We analyzed the Lyapunov exponents \citep{vogt2022lyapunov} of the inferred weights, which quantify the long-horizon behavior of the system, between preparation and execution. Values above zero indicate a tendency towards chaos, values below zero indicate stability, and values near zero are consistent with criticality or edge-of-chaos. We found that the transition between preparation and movement corresponded to an increase in the maximum Lyapunov exponent towards zero (Fig.~\ref{fig:embacc_monkey}.J \&K). This marginal stability reflects the motor system's trade-off between robustness and expressivity \citep{russo2018motor}.

\subsection{Uncovering fixed points in monkey motor cortex}
We lastly assessed the fixed point structure inferred by JEDI to test possible mechanisms governing the evolution of neural trajectories that underlie reaching movements in the motor cortex \citep{gallego2017neural}. Using the procedures described in the task-trained RNN experiment, we identified stable fixed points in the neural population activity (Fig.~\ref{fig:embacc_monkey}.L). These fixed points were clustered by reach direction, consistent with different attractor zones for different reaching conditions. Intriguingly, we found that neural trajectories identified by PCA ended at these stable fixed points; this end point corresponds to the end of the monkey's reach. 

The fixed points uncovered by JEDI, together with the eigenspectra and Lyapunov exponents, posits that neural dynamics for reaching are governed by marginally-stable trajectories towards stable attractors. This mechanistic insight highlights the strength of JEDI for analyzing neural data.

%% file: sections/related_work.tex
\section{Related Work}

RNNs are ubiquitous in computational neuroscience, with the aim to explain neural computations through the lens of underlying dynamical principles \citep{vyas2020computation}. Prior work highlighted the role of contexts and task variables in shaping neural computations \citep{yang2019task, driscoll2024flexible, costacurta2024structured, williams2025expressivity}. Typically, these studies use fixed (not learned) contextual inputs to probe cognitive and decision-making tasks. These works share an emphasis on contextual inputs with our work, but focus on training networks \textit{de novo} to learn specific tasks. They do not attempt to infer weight structures directly from experimentally-recorded neural activity, our primary goal.

Another common application of neural networks for neuroscience is latent state inference, including LFADS \citep{Pandarinath2018} and XFADS \citep{dowling2024exponential}.  CEBRA \citep{schneider2023learnable}, another latent state inference tool, learns latent embeddings jointly from neural and behavioral data, paralleling our use of context to align across conditions. Unlike these methods, ours aims to infer interpretable weight structure to reproduce the full time series of neural population recordings.

Existing dRNN methods focus on fixed dynamical systems. Methods such as CURBD \citep{Perich2020} and CORNN \citep{Dinc2023} directly learn from time series neural data. Others model neural activity at in the latent space with low-rank RNNs \citep{pals2024inferring,Valente2022}. Our method shares this low-rank assumption and its structural implications, but extends them in two ways: (1) we fit data at full neural resolution to enable the study of individual neural interactions; and (2) we explicitly link weight structure to learned context embeddings. Lastly, recent work on motor adaptation through low-tensor rank RNNs \citep{pellegrino2023low} captures trial-level variation in weights, which is conceptually similar to our context-dependent variations, though our work is more flexible, e.g. for modeling variations in behavioral tasks and even subjects.

Recent work has combined learned contexts with neural data modeling. Hierarchical state-space models (SSMs) \citep{vermani2024meta} and meta-learning frameworks \citep{cotler2023analyzing} integrate recordings across contexts, akin to our contextual embeddings. Hierarchical models for time series \citep{brenner2024learning} and Bayesian models of decision criteria \citep{vloeberghs2025bayesian} share our goal of extracting contextual structure \citep{kirchmeyer2022generalizing}. Relatedly, an SSM neural decoding architecture showed that task-specific embedding layers enable shared dynamics models ~\cite{ryoo2025generalizablerealtimeneuraldecoding}, though these lack the interpretability of JEDI. Lastly, context-informed dynamics models that generalize across physical systems \citet{nzoyem2025weight} share common goals with our work, as they explicitly study recurrent models in weight space. However, unlike their linearized architectures, our work applies to nonlinear, data-constrained RNNs. Ultimately, the above methods focus on using context learning to improve generalization and do not address our core aim: to mechanistically probe RNN weights.

%% file: sections/discussion.tex
\section{Discussion}


\textbf{Summary.} In this work, we introduced JEDI, a hierarchical hypernetwork model that generates low-rank autonomous RNNs to fit neural data. Our core objective was to confirm that contextual embeddings that parameterize recurrent weight matrices could yield interpretable and generalizable models of neural computations via dynamical systems. JEDI simultaneously reveals contextual relationships in neural activity between experimental conditions (e.g., behavioral tasks) and underlying dynamical mechanisms. 

We demonstrated that the low-rank hypernetworks used in JEDI simultaneously: (i) produced embeddings that reliably classified varying dynamical regimes and generalized to new data; and (ii) generated recurrent interaction matrices whose eigenvalue spectra aligned with known dynamical signatures in synthetic systems and revealed new motifs in real neural recordings. Our results indicate that context-conditioned weight generation uncovers invariant and equivariant structure in neural activity. Low-rank constraints act as an inductive bias that improves robustness and generalization. Importantly, spectral analysis of the learned interactions offers mechanistic insight to the dynamical rules governing computation. Prior low-rank RNN studies established that structured connectivity shapes computation. Our work extends this perspective by linking context embeddings to recurrent weight organization through hypernetworks, a first step towards bridging latent variable approaches with weight-space analyses. JEDI thus provides a scalable route to integrate heterogeneous datasets, generate hypotheses about neural dynamics, and probe conserved motifs across tasks.

Future work should pursue adaptive regularization schemes to relax rank specification, incorporate priors that capture biologically plausible connectivity when these variables are known \textit{a priori}, and extend the framework to settings with incomplete or multimodal observations. Applying the model to larger-scale neural datasets across tasks and species should reveal conserved organizational principles and further validate the capacity of hypernetwork-driven architectures to capture the structure of neural computation.

\textbf{Limitations}. The explicit specification of rank imposes a manual design choice; adaptive strategies such as nuclear norm regularization \citep{scarvelis2024nuclear} may alleviate this. The models remain susceptible to vanishing and exploding gradients, suggesting the need for stabilized architectures (e.g., CORNN \citep{rusch2020coupled}). Finally, the reliance on partial observations in neural recordings may introduce mechanistic mismatches, limiting generalizability to unobserved populations.



 


%% file: sections/appendix.tex
\section{Appendix}
\subsection{Additional details on Datasets}
\label{apd:first}

\subsubsection{Synthetic Memory Pro task}
\label{appx:ctd_task_des}
We followed the training procedure outlined in the Computation through Dynamics benchmark \citep{versteeg2025computation} and selected the MemoryPro task from the suite \citep{yang2019task}, as its fixed-point structure was previously characterized by \cite{driscoll2024flexible}. In this task, the RNN learned to respond in the same direction as a stimulus after a memory period by minimizing the squared loss between its 3-dimensional output and the target using backpropagation through time. The input space consisted of 18 dimensions, including a 1-d fixation signal, 2-d stimulus vectors encoding the circular variable $\theta$ as $A\sin(\theta)$ and $A\cos(\theta)$, and 15-d rule inputs that remained active throughout the trial to indicate the task type. We used a tanh activation function for the RNN and treated each task period—context, stimulus, memory, and response—as an autonomous dynamical system with a distinct set of fixed points induced by the piecewise constant inputs.We structured the experimental trials by initiating a 75-timestep context period where only fixation and rule inputs were active, followed by the stimulus period and a subsequent memory period that shared identical inputs with the context phase. When the fixation input dropped to zero during the response period, the network generated its directional output, which we monitored across 36 distinct stimulus angles. To ensure a robust dataset, we generated 20 variations for each angle through random sampling, resulting in a final collection of 2,880 trials. We then fit our model to these activations by initializing 2,880 context embeddings and training them together to capture the underlying task dynamics and specific patterns of activity induced by the trial phases.

\subsubsection{Motor Cortex Recordings}
\label{appx:motor_cortex_des}
Two monkeys were trained to perform a two-dimensional center-out reaching task using a planar that controlled a cursor on a screen. In each trial, the monkey moved to a central start position, waited through a variable delay, and then reached toward one of eight randomly selected targets arranged uniformly in a circle. A go cue signaled movement initiation, and successful trials required reaching the target within 1 second and holding for 0.5 seconds to receive a reward. We tested the applicability of the proposed approach on this data containing $c=160$ trials of $T=150$ timesteps of recordings from $N=117$ neurons.

\subsubsection{Multiple frequency Noisy Sine data}
\label{appx:mf_task_des}
Similar to multi-task synthetic teacher task framework, we generated chaotic sinusoidal datasets by driving chaotic teacher RNN with 10 increasing frequencies across 20 unique initializations of initial states. This procedure produced a comprehensive dataset of 200 trials characterized by complex, non-linear dynamics. We subsequently utilized these trajectories to train and evaluate both JEDI and JEDI-full, assessing their capacity to capture and reconstruct high-dimensional chaotic representations.

\subsection{Model Architecture and Training details}
\label{apd:architecture}
The latent dimension for all the methods is set to 16.

\subsubsection{VAE}
We used a two-layer Multi-Layer Perceptron(MLP) for both the encoder and decoder. The encoder transforms the flattened input of shape $(T \times features)$ into a latent vector of dimension 16 though hidden layer of 128 hidden units. The decoder mirrors this architecture, decoding from the latent space through a hidden layer of 128 units back to the input dimension.

The reconstruction loss is computed using the reparameterized latent vector, and the total loss includes a standard KL divergence term. During training, we set the batch size to 30 and trained the model for 1500 epochs, selecting the best model using validation loss. We used the Adam optimizer with an initial learning rate of  $10^{-3}$. Tanh activation was used in the decoder output. 

\subsubsection{RNN VAE}

For the RNN-based VAE model (RNNVAE), we used a sequence-to-sequence architecture built upon an LSTM encoder and decoder. The encoder comprises a single-layer LSTM with 32 hidden units, followed by a fully connected layer projecting into a latent space of size 16, where the mean and log-variance are estimated through parallel linear layers. The decoder reconstructs the sequence by transforming the latent vector through a linear projection and passing it through a LSTM layer. The final output is mapped back to the input dimensionality through a fully connected layer. Tanh activation was used in the decoder output. This model operated on a sequence length of 200 time steps.

Total loss was mean squared error between the decoder output and the ground truth trajectory and  a standard KL divergence term. During training, we set the batch size to 150 and trained the model for 5000 epochs, selecting used the validation loss. We used the Adam optimizer with  learning rate of  $10^{-2}$. Tanh activation was used in the decoder output. 

\subsection{Fixed point finding and visualization}

To evaluate the model's ability to capture ground-truth dynamical structures, we utilized a task-trained RNN as a benchmark. We trained this model on the MemoryPro task from the Computation through Dynamics repository \citep{versteeg2025computation} and simulated trials across four distinct phases: context, stimulus, memory, and response. We qualitatively assessed the true and inferred dynamics by analyzing their fixed-point structures. Fixed points represent regions in the state space where dynamics are slow enough to permit linear approximation, revealing the system's local stability and behavior. To identify these points, we located coordinates in the hidden state space ($z \in \mathbb{R}^{d_z}$) that minimized the system's kinetic energy, defined as $q \approx \|\Delta h\|^2$. We performed this analysis using a modified version of the fixed-point finding toolkit developed by \cite{golub2018fixedpointfinder}, adapting it for autonomous RNNs. For visualization, we applied Principal Component Analysis (PCA) to the hidden state trajectories and projected the identified fixed points into this low-dimensional subspace. We used speed tolerance of $1e-15$ for MemoryPro task and $1e-9$ for motor cortex recordings.

\subsection{Details of Lyapunov Exponents calculation}
To calculate the Lyapunov exponents, we followed algorithm proposed by \cite{engelken2023lyapunov} the algorithm tracks how tiny perturbations to the network's state grow or shrink over time by evolving a set of test vectors alongside the main simulation. The algorithm uses jacobian matrix to update these vectors, measuring how the network activations amplify or dampen small differences.  By averaging the logarithms of the stretching factors (the  values from the decomposition) over the entire simulation, we can determine the final exponents. While a positive exponent confirms that the network is chaotic, stable dynamics is characterized by negative exponent. 
\subsection{Supplementary Experiments and Details}

\subsubsection{Model reconstructions}
\begin{figure}[!ht]
\includegraphics[scale=1.0, center]{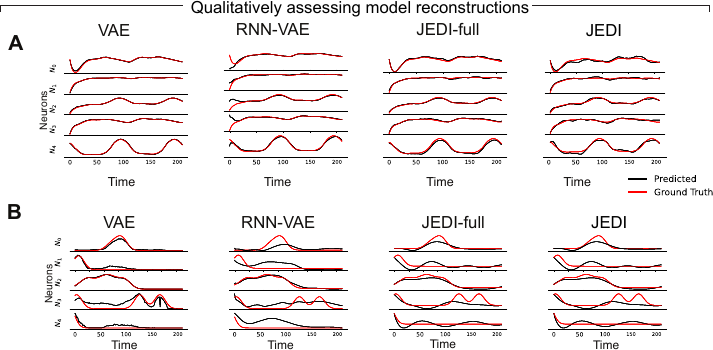}
\caption{Reconstructions of neural trajectories from various trained models. A) Reconstructions of Synthetic Data. B) Reconstructions of the Monkey cortex data.
}
\label{fig:appen_recon}
\end{figure}

\subsubsection{Hyperparameter Sweep}

\begin{figure}[H]
\includegraphics[scale=1.0, center]{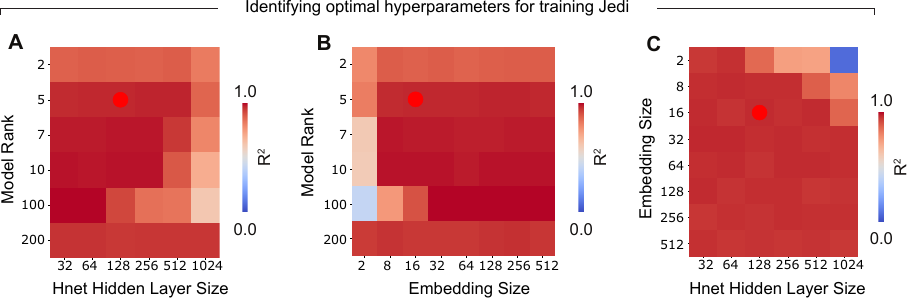}
\caption{Jedi performance on the synthetic data generated by the teacher RNN. The heatmaps show accuracy of reconstruction as measured by the coefficient of determination $R^2$ score. The red  circular dot highlights the hyperparameter that was chosen for experiments. A) $R^2$ heatmap for various model rank  vs Hypernetwork hidden layer size. B) $R^2$ heatmap for various Model rank vs Embedding size of the hypernetworks and c) $R^2$ heatmap for various embedding size vs Hidden size  of the hypernetwork. }
\label{fig:appen_hyperparameter}
\end{figure}

\subsubsection{Effect of data rank on performance of Jedi}

\begin{figure}[!ht]
\includegraphics[scale=0.3, center]{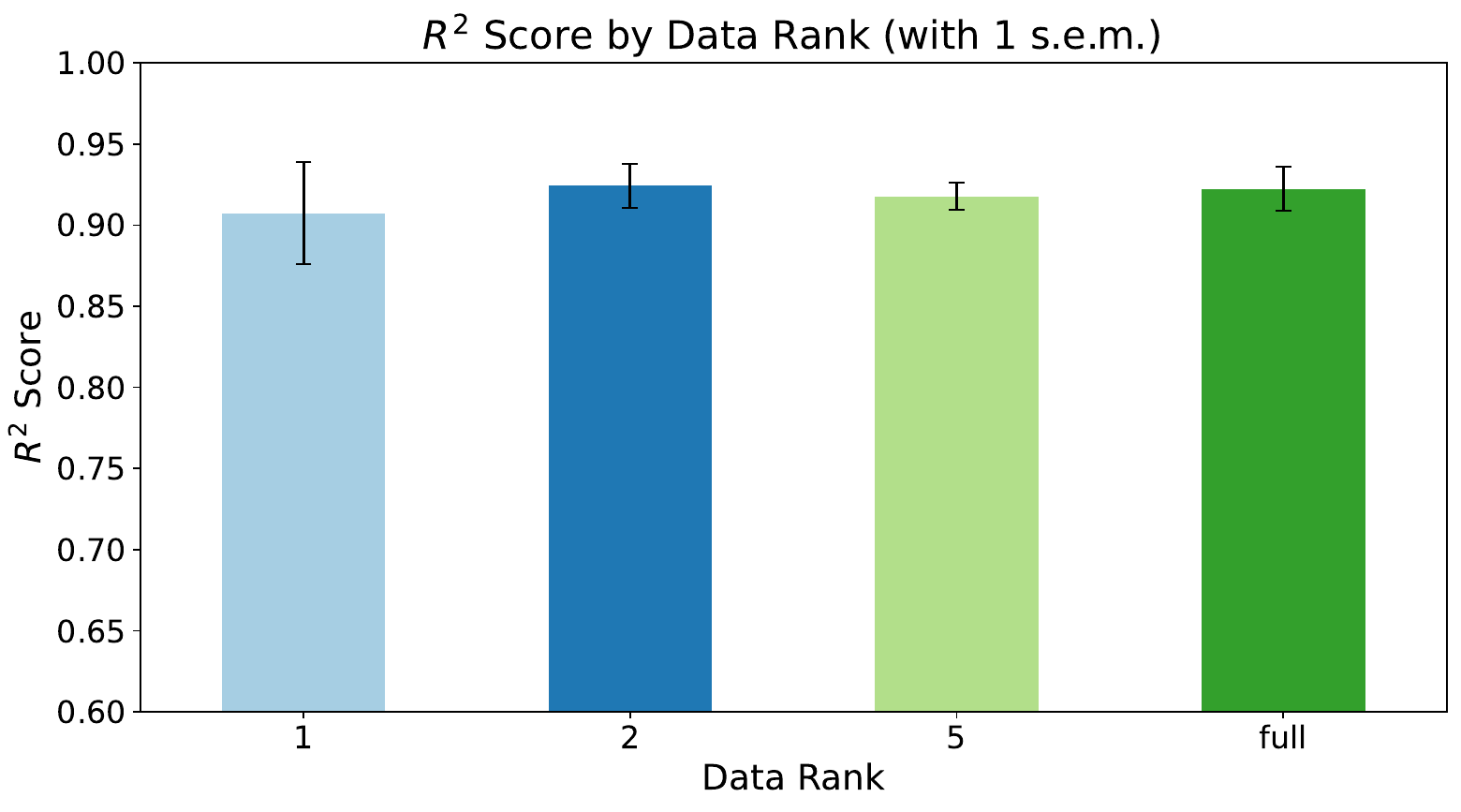}
\caption{Performance for Jedi on the multi-task synthetic data generated by the teacher RNN with different rank of connectivity matrix $J$. The bar plot shows accuracy of reconstruction as measured by coefficient of determination $R^2$ score (with 1 s.e.m).  Jedi was able to fit well on varied data ranks with the generated weights set to rank=5.}
\label{fig:appen_data_rank}
\end{figure}

\subsection{Additional results on spectral analysis}


\begin{figure}[!ht]
\includegraphics[scale=1.0, center]{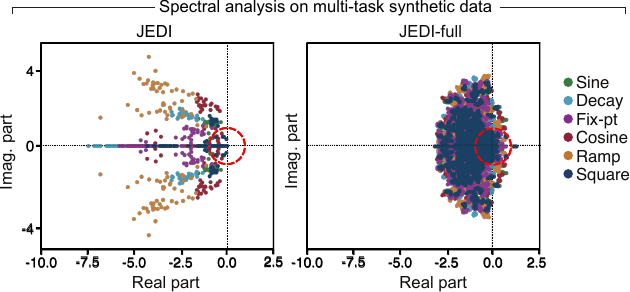}
\caption{Eigenvalue spectra of learned recurrent weights on multi-task synthetic data. JEDI learned consistent structure across tasks, while JEDI-full exhibited a dense, isotropic cloud without task separation.}
\label{fig:appen_full_synth}
\end{figure}

\begin{figure}[!ht]
\includegraphics[scale=1.0, center]{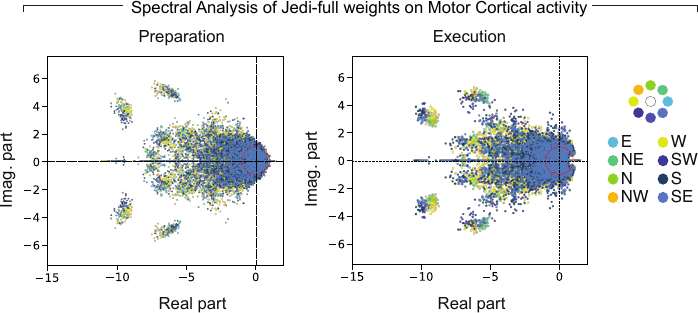}
\caption{Eigenvalue spectra of Jedi-full weights on monkey motor cortical activity. JEDI-full exhibited a dense, isotropic cloud without task/direction separation.}
\label{fig:appen_full_motor}
\end{figure}

\begin{figure}[!ht]
\includegraphics[scale=1.0, center]{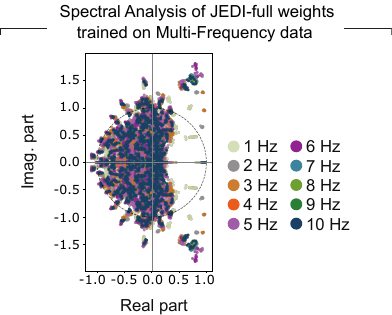}
\caption{Eigenvalue spectra of learned recurrent weights on multiple frequency dataset. The legend indicates the specific sine frequencies used to train the models, JEDI-full consistently exhibits a dense, isotropic spectral cloud. This distribution lacks any discernible structure or cluster formation corresponding to the underlying task frequencies.}
\label{fig:appen_full_multi_fre}
\end{figure}




\subsection{Additional results on monkey reaching data (Preparation phase)}
\begin{figure}[H]
\includegraphics[scale=1.0, center]{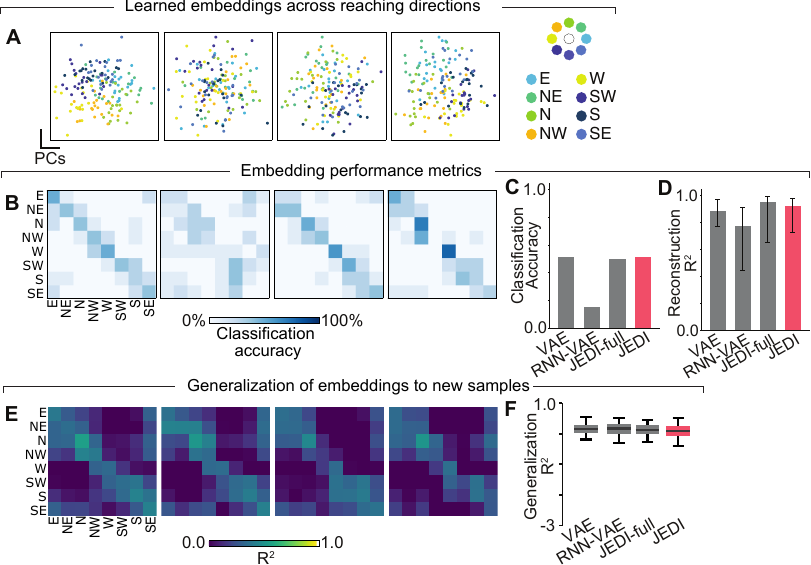}
\caption{Assessing embedding quality on monkey reaching data during preparation.
a) 2D PCA visualization of embeddings from different models. Each point represents a sample trial, color-coded by reach direction.
b) Confusion matrices of direction classification accuracies. Rows correspond to the true direction, and columns to the predicted direction.
c) Confusion matrices showing generalization performance across reach directions. Each cell indicates the $R^2$ score when decoding one direction (column) using mean embeddings of the other(row).}
\label{fig:add_prep}
\end{figure}

\subsection{Additional results on generalization}
\begin{figure}[H]
\includegraphics[scale=1.0, center]{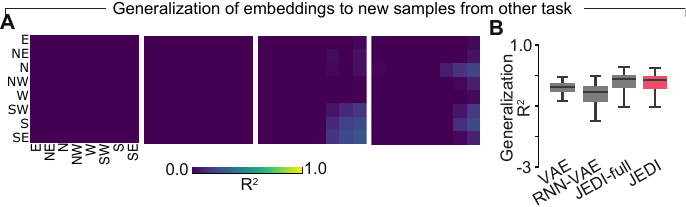}
\caption{Assessing embedding quality on monkey reaching data during preparation.
a) 2D PCA visualization of embeddings from different models. Each point represents a sample trial, color-coded by reach direction.
b) Confusion matrices of direction classification accuracies. Rows correspond to the true direction, and columns to the predicted direction.
c) Confusion matrices showing generalization performance across reach directions. Each cell indicates the $R^2$ score when decoding one direction (column) using mean embeddings of the other(row).}
\label{fig:add_gen}
\end{figure}

\subsection{Additional fixed points results}

\begin{figure}[H]
\includegraphics[scale=1.0, center]{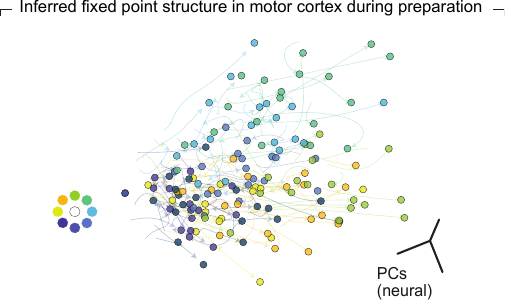}
\caption{Stable fixed points during
preparation inferred by JEDI colored by reach directions. Single-trial neural trajectories for each corresponding reach is also plotted.}
\label{fig:fp_prep}
\end{figure}

\begin{figure}[H]
\includegraphics[scale=1.0, center]{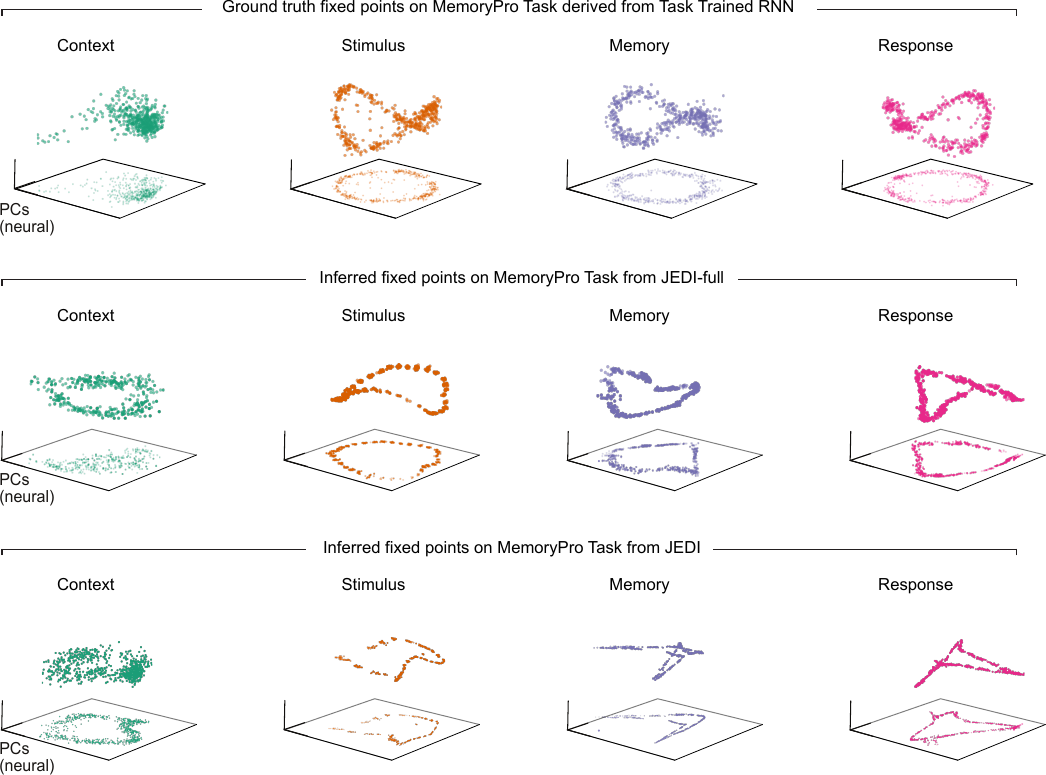}
\caption{3d PCA projection of fixed points from Jedi across 4 periods on MemoryPro task. The fixed points structure closely resembled the structure of task-trained RNN.}
\label{fig:ft_hnet}
\end{figure}

\begin{figure}[!ht]
\includegraphics[scale=1.0, center]{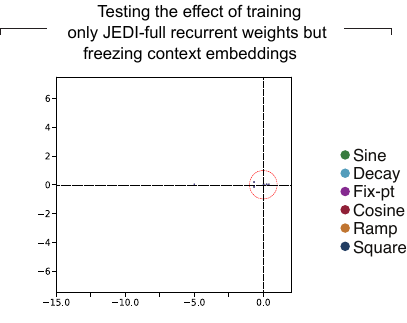}
\caption{When the context embedding remains frozen during training, the eigenvalue spectra of the learned recurrent weights fail to capture task-specific signatures. Consequently, joint optimization of both contexts and weights is essential to develop meaningful, trial-specific representations.}
\label{fig:appen_no_context}
\end{figure}

\subsection{Compute Resources}
We list below the compute resources used per experiment:
\begin{enumerate}
\item Multi-Task Teacher experiment: Results were computed on external cluster equipped with Nvidia L40S GPUs. Training wall clock times was 6hrs. Inference converges in a few seconds.
\item Multi-Frequency Since experiment: Results were computed on external cluster equipped with Nvidia L40S GPUs. Training wall clock times was 6hrs. Inference converges in a few seconds. 
\item Task trained RNN experiment: Results were computed on external cluster equipped with Nvidia L40S GPUs. Training wall clock times was 1.5-2 days. Inference converges in a few seconds.
\item Monkey reaching experiment : Results were computed on external cluster equipped with Nvidia L40S GPUs. Training wall clock times was 6hrs. Inference converges in a few seconds.
\end{enumerate}